\documentclass[12pt,letterpaper,notitlepage]{article}
\pdfoutput=1
\usepackage{float} 
\usepackage{amsmath} 
\usepackage{graphicx} 
\usepackage{cite}

\newcommand{\starttext} {\onecolumn}

\floatstyle{plain}
\newfloat{suppFigure}{!h}{lop}
\floatname{suppFigure}{Supplementary Figure}

\newfloat{suppVideo}{!h}{lop}
\floatname{suppVideo}{Supplementary Video}

\newcommand{\um}     {\,\mathrm{\mu m}}
\newcommand{\mm}   {\,\mathrm{mm}}
\newcommand{\ms}     {\,\mathrm{ms}}
\newcommand{\us}     {\,\mathrm{\mu s}}
\newcommand{\Hz}    {\,\mathrm{Hz}}
\newcommand{\pix}    {\,\mathrm{pixels}}

\newcommand{\degC}   {^\circ\mathrm{C}}

\newcommand{\mW}    {\,\mathrm{mW}}
\newcommand{\mL}  {\,\mathrm{mL}}
\newcommand{\hours}   {\,\mathrm{h}}
\newcommand{\cm}  {\,\mathrm{cm}}

\title{Phase gradient microscopy in thick tissue with oblique back-illumination}

\author{Tim N Ford, Kengyeh K Chu \& Jerome Mertz\\
	Department of Biomedical Engineering\\
	Boston University\\
	Boston, MA 02215\\
	\texttt{timford@bu.edu,jmertz@bu.edu}}

\date{\today}

\begin{document}

\maketitle

\begin{abstract}
Phase contrast techniques, such as differential interference contrast (DIC) microscopy, are widely used to provide morphological images of unstained biological samples. The trans-illumination geometry required for these techniques has restricted their application to thin samples. We introduce oblique back-illumination microscopy (OBM), a method of collecting \emph{en face} phase gradient images of thick scattering samples, enabling near video-rate \emph{in vivo} phase imaging with a miniaturized probe suitable for endoscopy.
\end{abstract}

\starttext
Phase contrast microscopy techniques are widely used in biological research because they can provide high resolution images of unlabeled samples even when these are nearly transparent. For example, differential interference contrast (DIC), which reveals lateral phase gradients, is one of the more popular techniques because it provides apparent 3D sample relief using a standard microscope equipped with a lamp and a camera \cite{Nomarski1955}. Even simpler techniques based on oblique illumination can also be used with standard microscopes and provide similar imaging as DIC \cite{Saylor1935, Axelrod1981, Yi2006, Mehta2009}. However, in order to reveal lateral phase gradients, all these techniques must be operated in trans-illumination configurations, limiting their use to thin samples. Because one is often constrained to working instead with thick samples (e.g. endoscopy or \emph{in vivo} applications), there is a clear need for a method that can provide DIC-like imaging in thick samples. We introduce such a method here.

\begin{figure}[!htp]
\begin{center}
\includegraphics{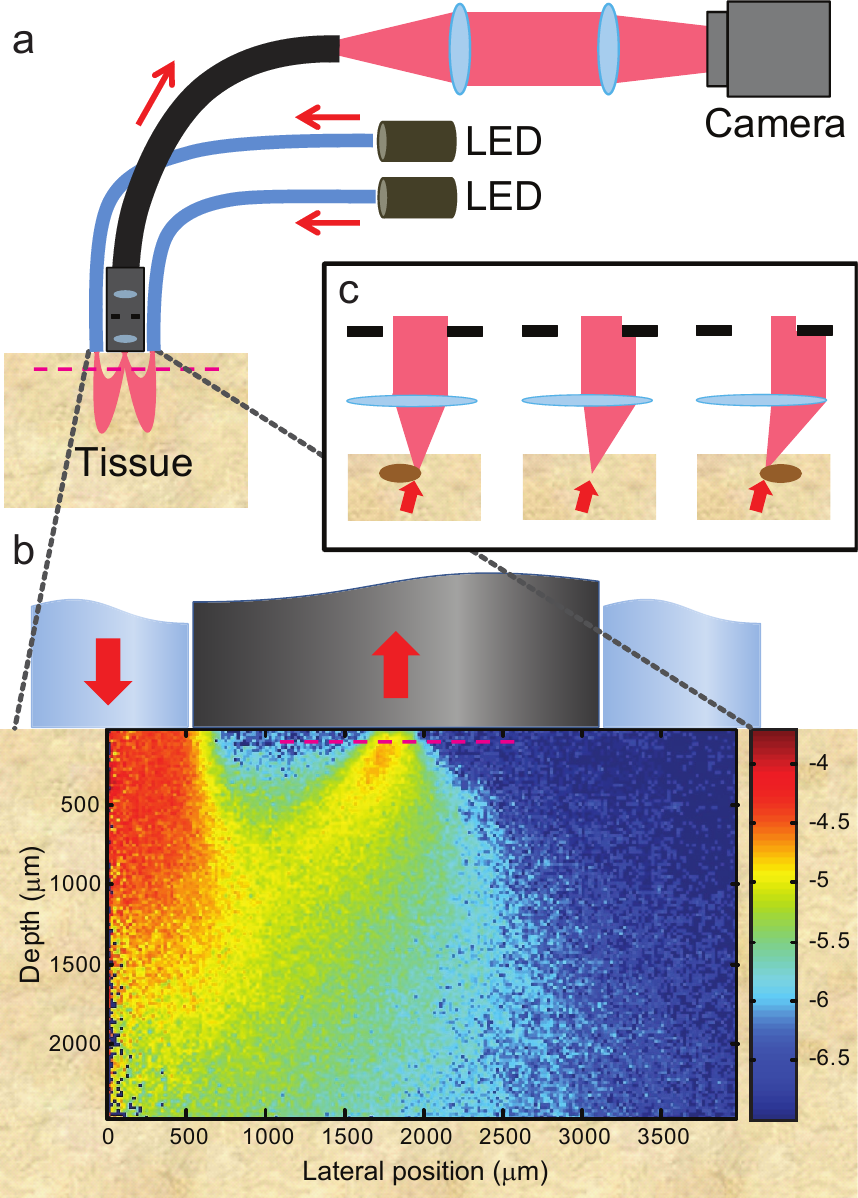}
\caption{An OBM setup with a contact-mode endomicroscope probe. (\textbf{a}) Illumination from two LEDs is sequentially projected into a thick sample by diametrically opposed optical fibers attached to the probe housing. Multiple scattering in the sample redirects the light so that it trans-illuminates the focal plane of the probe micro-objective (magenta dashed line). An image from the focal plane is then relayed by a flexible fiber bundle and projected onto a digital camera.  (\textbf{b}) Close-up of the probe distal end, onto which is superposed a density map obtained by Monte Carlo simulation of the light energy in the sample that was injected by a single fiber (from left) and collected by the micro-objective (Online Methods). Note the obliqueness of the light distribution through the focal plane. (\textbf{c}) Oblique trans-illumination is partially blocked by the micro-objective back aperture. Index of refraction variations at the focal plane refract the light causing changes in the image intensity that are proportional to the slope of the variations, thus leading to phase gradient contrast.
\label{fig:setup}}
\end{center}
\end{figure}

Our technique is called oblique back-illumination microscopy (OBM). As its name suggests, this technique is based on a similar principle as oblique illumination microscopy, with the notable difference that illumination and detection both occur from the same side of the sample (i.e. in a reflection geometry), allowing its application to samples of arbitrary thickness. Unlabeled samples can, of course, be imaged with microscopes based on direct light reflection. The most successful of these for tissue imaging is optical coherence tomography (OCT), which, like OBM, can also be operated in a widefield \emph{en face} configuration \cite{Boccara2009}. However, microscopes based on light reflection intrinsically reveal only sample structure that varies rapidly in the axial direction, such as sharp interfaces or particles much smaller than the illumination wavelength \cite{Mertz2009}. In contrast, microscopes based on light transmission are not subject to this constraint, and can reveal even slowly varying sample structure in the lateral direction, thus providing images of subtle sample features impossible to see in reflection mode.  An important characteristic of OBM is that, even though it is configured in a reflection geometry, it is, in fact, a transmission microscope in disguise. In effect, OBM uses multiple scattering in tissue to convert epi-illumination into trans-illumination. Because the illumination source is offset from the detection optical axis, the trans-illumination is oblique (\textbf{Fig.~\ref{fig:setup}}). While illumination obliquity directly leads to phase gradient contrast, image intensity is also influenced by sample absorption. The use of two off-axis sources diametrically opposed to one another permits the acquisition of two raw images with similar absorption contrast but with phase gradient contrast of opposing sign (\textbf{Supplementary Fig.~\ref{fig:rawAddSub}}). The subtraction of these raw images enhances phase gradient contrast while canceling absorption contrast; addition of the raw images has the opposite effect, revealing only absorption contrast while canceling phase gradient contrast. By this method, the sequential acquisition of two raw images using alternating illumination sources decouples absorption and phase gradient contrast.

We present results obtained from a miniaturized OBM built using a flexible endomicroscope probe comprising a distal micro-objective and an imaging fiber bundle comprising 30,000 fiber cores (Online Methods). Illumination from two independently controlled light emitting diodes (LEDs) was delivered to the sample via two optical fibers attached opposite one another to the micro-objective housing. The endomicroscope probe was designed to operate in contact mode, meaning that light reflected directly from the sample surface is not collected by the micro-objective. Instead, the micro-objective only collects light that has been multiply scattered in the sample and redirected upwards through the focal plane, located here at a depth of $60\um$ (the working distance of the micro-objective). Unless stated otherwise, all images presented here are individual frames from movies acquired and displayed at a net rate of $17.5\Hz$ using a double-shutter camera that reads images pairwise. The exposure time per raw image was typically 1 to $5\ms$.

\begin{figure}[!htp]
\begin{center}
\includegraphics{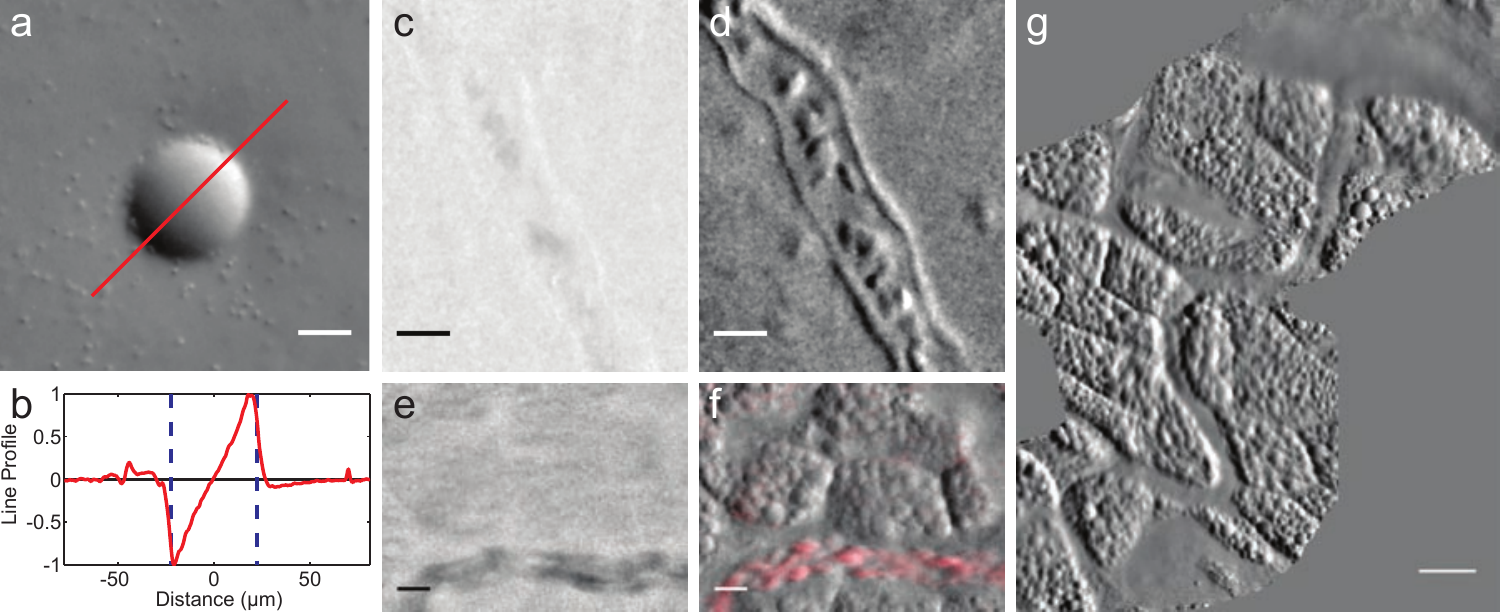}
\caption{Demonstration of OBM in a tissue phantom and \emph{in vivo}. (\textbf{a}) Phase gradient image of a $45\um$ polystyrene bead in scattering medium ($2\um$ beads in agarose, \textbf{Supplementary Video~\ref{fig:beadVid}}). Note $2\um$ beads are readily visible. (\textbf{b}) Corresponding phase gradient profile. (\textbf{c-f}) Simultaneously acquired absorption (\textbf{c,e}) and phase gradient (\textbf{d,f}) images of the CAM vascular system of day 11 chick embryo. Individual red blood cells (RBCs) and vessel walls are clearly visible (\textbf{Supplementary Video~\ref{fig:capillaryVid}}). (\textbf{f}) Moving RBCs are highlighted in red using a sliding 3-frame temporal variance filter (\textbf{Supplementary Video~\ref{fig:red1XVid}}). (\textbf{g}) A CAM vasculature mosaic reconstructed from \textbf{Supplementary Videos~\ref{fig:absphaVid}} and \textbf{\ref{fig:mosaicVid}}. Scale bars are (\textbf{a},\textbf{c-f}) $20\um$, and (\textbf{g}) $50\um$.
\label{fig:chick}}
\end{center}
\end{figure}

\textbf{Figure~\ref{fig:chick}} shows a phase gradient image of a $45\um$ polystyrene bead embedded in a scattering tissue phantom consisting of $2\um$ beads mixed in agarose. The phase gradient induced by the bead was observed to be approximately linear. Note that since the phase gradient image is derived from a difference of raw images, it contains both positive and negative values (zero is represented by an intermediate gray level). Because phase gradients must, by definition, arise from apparent sample structure, they must also arise from the vicinity of the focal plane (objects out of focus are blurred and exhibit little structure). Thus, phase gradient imaging exhibits apparent out-of-focus background rejection. This is manifest in \textbf{Supplementary Video~\ref{fig:beadVid}}, in which only the in-focus $2\um$ beads are visible.
 
\textbf{Figure~\ref{fig:chick}} also shows simultaneously acquired absorption and phase gradient images of the chorioallantoic membrane (CAM) in a day 11 chick embryo \emph{in vivo} and \emph{in ovo}. Note the low contrast of the absorption images compared to the phase gradient images. Because the absorption images are derived from the sum of raw images, their values are positive definite (zero is black), and they do not exhibit out-of-focus background rejection (such imaging is similar to orthogonal polarization spectral imaging \cite{Groner1999}). Compared to images obtained from OCT, OBM images are speckle-free. Moreover, owing to the large photon fluxes involved, they are also relatively shot-noise free. Intensity noise arises, in our case, mostly from inhomogeneous image sampling due to an uneven distribution of fiber cores in the imaging fiber bundle (Online Methods).

\begin{figure}[!htp]
\begin{center}
\includegraphics{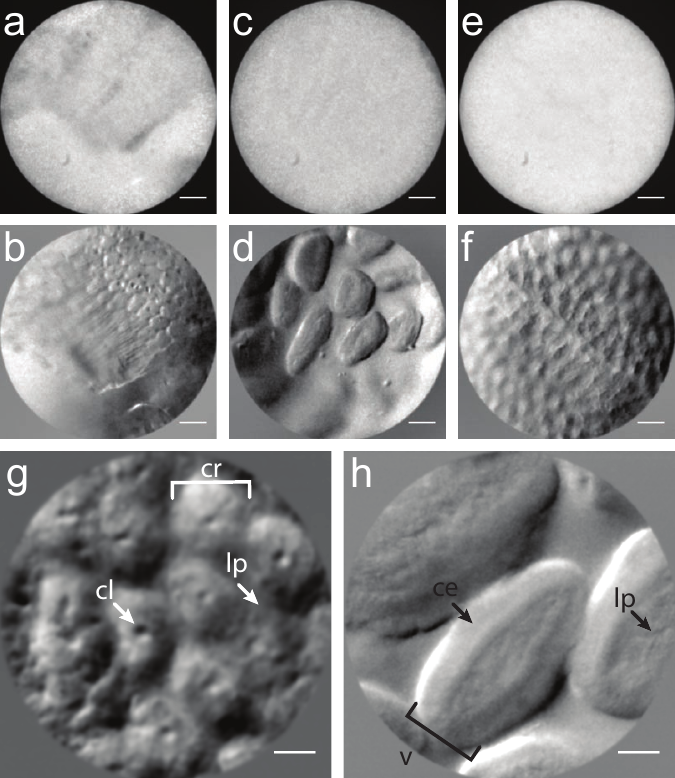}
\caption{Demonstration of OBM in excised mouse intestinal epithelium. (\textbf{a-f}) Simultaneously acquired amplitude (\textbf{a,c,e}) and phase gradient (\textbf{b,d,f}) images taken with a $1\times$ micro-objective (field of view (FOV) $600\um$). (\textbf{g,h}) Higher magnification phase gradient images of the epithelium in the distal colon (\textbf{g}) and small intestine (\textbf{h}) were taken with a $2.5\times$ micro-objective (FOV $240\um$). Crypts of Lieberk\"{u}hn (cr), crypt lumens (cl) and lamina propia (lp) are indicated with arrows (\textbf{g}, \textbf{Supplementary Video~\ref{fig:colonVid}}); ileal villi (v), columnar epithelium (ce) and lamina propia (lp) are indicated with arrows (\textbf{h}, \textbf{Supplementary Video~\ref{fig:ileumVid}}). Scale bars are (\textbf{a-f}) $75\um$, and (\textbf{g,h}) $30\um$.
\label{fig:mouse}}
\end{center}
\end{figure}

Finally, \textbf{Figure~\ref{fig:mouse}} shows simultaneously acquired absorption and phase gradient images of excised, unstained mouse intestinal epithelium. The absorption images are essentially featureless in this case, whereas the phase gradient images are information rich. For example, crypts of Lieberk\"{u}hn and ileal villi are readily visible in \textbf{Figures~\ref{fig:mouse}g} (\textbf{Supplementary Video~\ref{fig:colonVid}}) and \textbf{\ref{fig:mouse}h} (\textbf{Supplementary Video~\ref{fig:ileumVid}}), respectively, illustrating the potential of OBM for \emph{in situ} histopathology and ``optical biopsies''.

In summary, we have presented an apparatus that, to our knowledge, is the first to provide sub-surface, DIC-like (i.e. transmission-like) phase gradient imaging from thick scattering tissue in a reflection geometry. The apparatus is simple, fast, robust, and inexpensive, making it broadly appealing to biological and clinical researchers alike.

\section*{Acknowledgments}
We thank S.~Singh and J.~Ritt for supplying the mouse gastrointestinal samples. We thank K.~Calabro for helping develop the Monte Carlo simulation code. We thank all the members of the Biomicroscopy Lab for their helpful conversations and careful review of this manuscript. This work was supported by an NIH grant R01-EB010059.

\section*{Author Contributions}
T.N.F., K.K.C. and J.M. conceived and developed the technique. T.N.F. built the setup and acquired the data. T.N.F. and J.M. wrote the manuscript. J.M. supervised the project.

\section*{Competing Financial Interests}
The authors declare no competing financial interests.

\section*{ONLINE METHODS}
\subsection*{Hardware setup}
White light from two LEDs (Luxeon Star MR-WC310-20s) was coupled into optical fibers (Thorlabs BFL48-1000; 0.48~NA; $1000\um$ core) using aspheric condenser lenses (Thorlabs ACL5040-A). Illumination light was launched by the fibers into the sample ($25\mW$ per channel at the fiber output), where it was redirected through the focal plane by multiple scattering and collected by a micro-objective (Mauna-Kea Technologies; $2.6\mm$ diameter; $1\times$ or $2.5\times$ magnification; $60\um$ working distance; water-immersion; 0.8~NA) coupled to a coherent imaging fiber bundle (30,000~cores;  $600\um$ active area). The separation distance between the fiber and the micro-objective probe was approximately $1.8\mm$. The proximal face of the fiber bundle was imaged with standard microscope optics (Olympus Plan $10\times$ 0.48~NA air objective, Linos AC $f=200\mm$ tube lens; $4f$ configuration) and recorded with a digital camera (PCO Pixelfly USB; 14-bit; $2\times2$ binning; 35~fps; $1-5\ms$ exposure time per illumination direction). The camera was operated in double shutter mode to reduce the inter-frame delay between exposures ($200\us$), minimizing motion artifacts \cite{Ford2012}. Illumination power delivered by the left and right optical fibers was triggered (Thorlabs LEDD1B) to overlap with the first and second frame in the each image pair, respectively. Frame rate was limited by the camera readout time. Image acquisition and display was performed using custom written software (National Instruments LabVIEW~11.0). Illumination gating and camera exposure were synchronously controlled using a data acquisition card (National Instruments PCI-6221).

\subsection*{Image processing}
A preprocessing routine described previously \cite{Ford2012} was first used to correct for the quasi-periodic sampling pattern imparted by the fiber bundle cores. Each raw image was then normalized by its respective low-pass filtered version (Gaussian filter kernel with $\sigma=80\pix$) to correct for non-uniform illumination profiles and thus ``flatten'' the images. The two normalized images were then either added or subtracted to produce absorption-only or phase gradient-only images, respectively (\textbf{Supplementary Fig.~\ref{fig:rawAddSub}}). Image processing was performed with a graphics processing unit (NVIDIA GTX280) using custom-written software written in CUDA-C \cite{CUDA}.

\subsection*{Monte Carlo simulations}
CUDAMCML \cite{Alerstam2009}, a modification of MCML \cite{Wang1995} enabling execution on graphics processing units (GPUs), was used to perform the simulations. CUDAMCML was further modified to execute on a cluster of CUDA-enabled workstations \cite{Calabro2012}. A semi-infinite slab geometry was modeled with tissue optical parameters $n_\mathrm{tissue}=1.37$, $l_s=150\um$, $l_s^*=3000\um$ and $g=0.95$ ($n$ is index of refraction, $l_s$ and $l_s^*$ are the scattering and transport mean free path lengths, respectively, and $g=1-l_s/l_s^*$ is the anisotropy factor). Illumination fiber parameters were $n_\mathrm{fiber}=1.37$, $\mathrm{diameter}=1000\um$ and numerical aperture, $\mathrm{NA}=0.48$. Micro-objective probe parameters were $n_\mathrm{probe}=1.37$, $\mathrm{diameter}=240\um$ and $\mathrm{NA}=0.8$. Fiber-probe separation was $d=1818\um$. A Henyey-Greenstein phase function was used to characterize photon scattering events \cite{Henyey1941}. $10^8$~photons were processed to estimate the distribution of exit angles of the detected photons as a function of fiber-probe separation (\textbf{Supplementary Fig.~\ref{fig:angularDist}}). Both the total detected intensity and median exit angle were observed to decrease with increasing fiber-probe separation. $10^5$~photons were processed to estimate photon path density as a function of lateral position and depth, revealing the so-called photon banana (\textbf{Fig.~\ref{fig:setup}b}).

\subsection*{Tissue phantom preparation}
The scattering tissue phantom was prepared by heating a $30\mL$ solution of $2\%$~(w/v) agarose (Sigma A5093-100G), $5\%\,2\um$ diameter polystyrene beads (Polysciences 19814-15), and $0.1\%\,45\um$ diameter polystyrene beads (Polysciences 07314-5) in $\mathrm{H_2O}$ to $75\degC$ on a hotplate, followed by pouring the mixture into a $60\mm\times15\mm$ cell culture dish (Corning 430166). The phantom was covered with paraffin film and left to cool to room temperature before imaging. The optical properties of the bulk medium were $l_s=74\um$, $l_s^*=1040\um$ and $g=0.93$, as estimated using Mie theory. The indices of refraction of hydrated agarose gel and polystyrene beads were $n=1.35$ and $n=1.59$, respectively \cite{Pogue2006}. Imaging was performed through water.

\subsection*{Chick embryo preparation}
Fertilized \emph{Gallus gallus} eggs (Carolina Biological Supply Co. 139290) were stored in an incubator at $37\degC$ and $50\%$ humidity, being turned every $7\hours$ to prevent fusion of the chorioallantoic membrane (CAM) with the shell membrane. Imaging was performed at embryonic day 11. A $1\cm$ diameter region of the shell and shell membrane was removed exposing the embryo and CAM. A layer of $37\degC$ saline was dripped over the preparation before imaging \emph{in ovo} with the OBM probe. Following imaging, the embryos were euthanized by hypothermia by storing the eggs at $-15\degC$.

\subsection*{Mouse tissue preparation}
Six week old C57~black~6 mice were euthanized by $\mathrm{CO_2}$ inhalation and the gastrointestinal tract was immediately excised and washed with $4\%$ paraformaldehyde. The colon and small intestine were cut longitudinally, unrolled and cleared of fecal matter. The preparations were stored in $4\%$ paraformaldehyde for several days. Before \emph{ex vivo} imaging, the tissues were pinned to a silicone elastomer slab (Sylgard\textregistered~184, Corning) to expose the apical surfaces. Residual mucus and fecal matter were gently washed away with saline before imaging. The animals used in this study were treated in accordance with the guidelines of the Institutional Animal Care and Use Committee of Boston University.

\newpage

\newpage

\Large\noindent\textbf{Supplementary Information}
\newline
\LARGE\noindent\textbf{Phase gradient microscopy in thick tissue with oblique back-illumination}
\newline
\large\noindent\textbf{Tim N Ford, Kengyeh K Chu \& Jerome Mertz}
\normalsize
\newline
\begin{center}
\begin{tabular}{ | l | l | }
	\hline
	Supplementary Figure 1 & Comparison of added versus subtracted raw OBM images\\ \hline
	Supplementary Figure 2 & Photon exit angle distribution estimated with Monte Carlo simulation\\ \hline
	Supplementary Video 1 & Manual focusing through scattering tissue phantom\\ \hline
	Supplementary Video 2 & CAM vasculature and demonstration of axial resolution\\ \hline
	Supplementary Video 3 & CAM vasculature with moving RBCs highlighted in red\\ \hline
	Supplementary Video 4 & Comparison of absorption versus phase gradient images in CAM\\ \hline
	Supplementary Video 5 & Phase gradient mosaic of CAM vasculature\\ \hline
	Supplementary Video 6 & Morphological features of mouse distal colon\\ \hline
	Supplementary Video 7 & Morphological features of mouse small intestine\\ \hline
\end{tabular}
\end{center}

\begin{suppFigure}
\begin{center}
\includegraphics{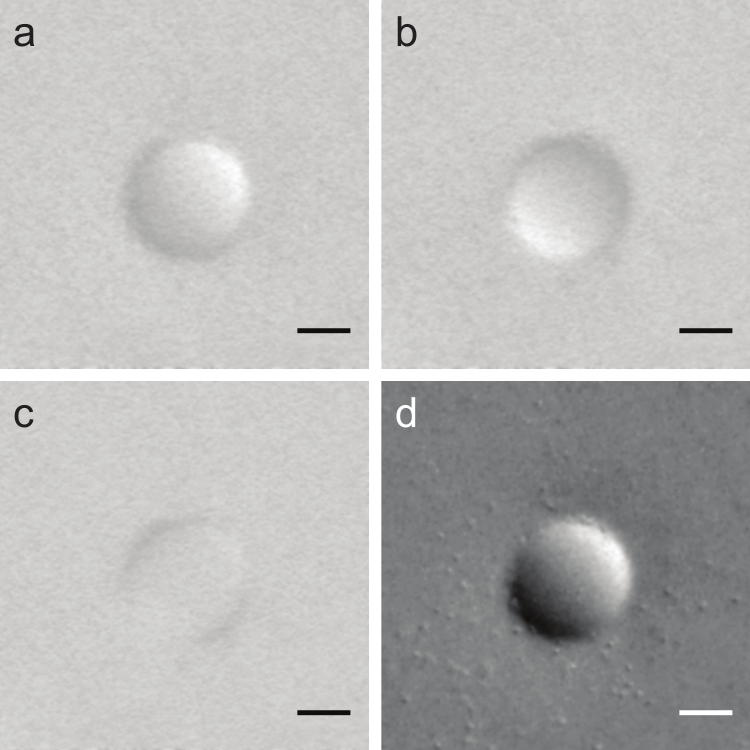}
\caption{$45\um$ polystyrene bead suspended in scattering tissue phantom. (\textbf{a,b}) raw images under oblique back-illumination from two opposing directions. (\textbf{c}) Addition of (\textbf{a}) and (\textbf{b}) cancels phase gradient contrast and emphasizes absorption. (\textbf{d}) Subtraction of (\textbf{a}) and (\textbf{b}) cancels absorption contrast and emphasizes phase gradients. The $2\um$ beads used to build the tissue phantom are readily visible only when they are in focus. Scale bars $20\um$.
\label{fig:rawAddSub}}
\end{center}
\end{suppFigure}

\begin{suppVideo}
\begin{center}
\includegraphics{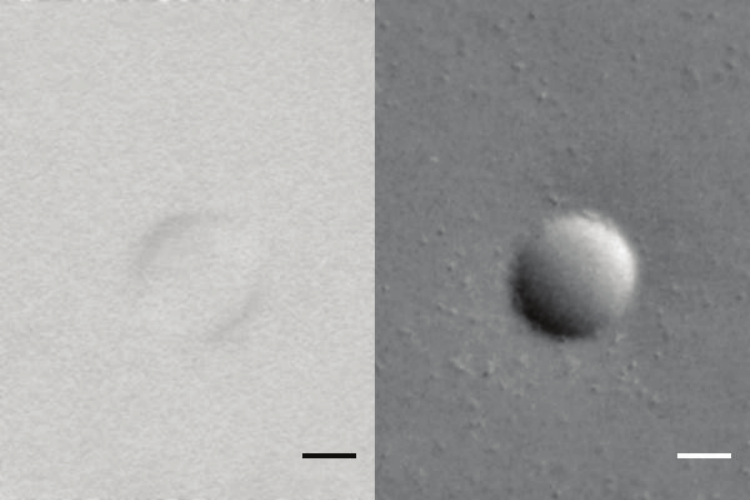}
\caption{OBM exhibits apparent axial resolution, as is demonstrated by focusing through suspended polystyrene beads. Scale bars $20\um$, imaging speed $5\Hz$.
\label{fig:beadVid}}
\end{center}
\end{suppVideo}

\begin{suppFigure}
\begin{center}
\includegraphics{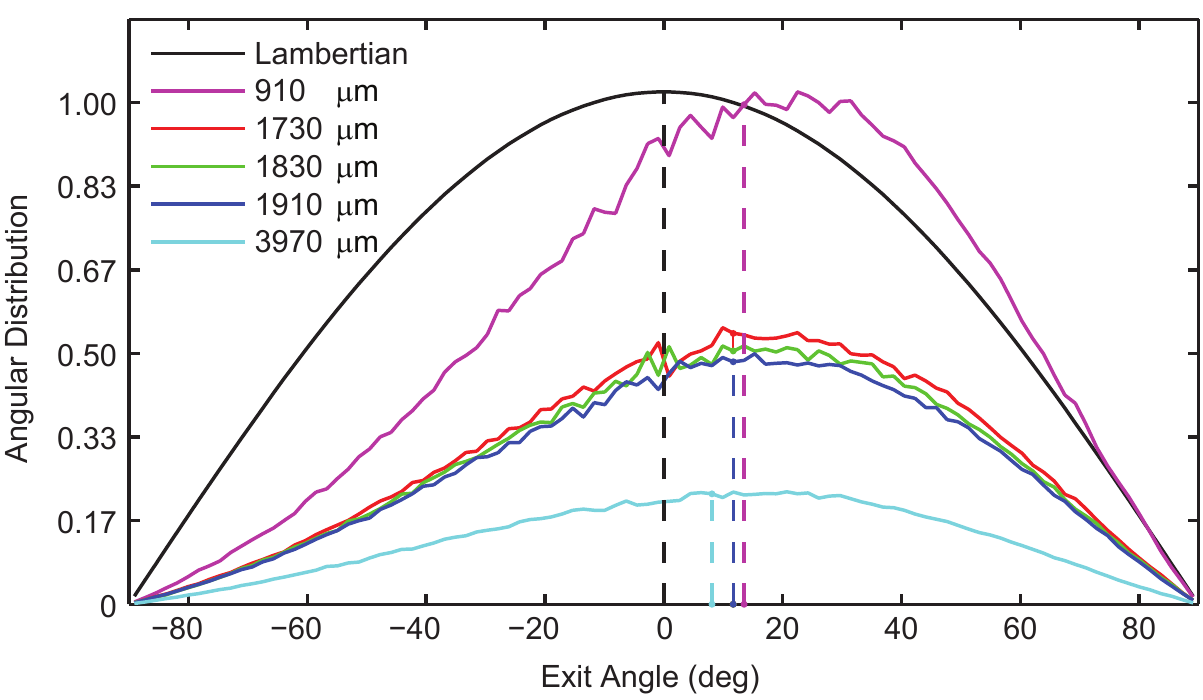}
\caption{Monte Carlo simulations estimate photon exit angle distributions at different fiber-detector separations. The exit angle corresponds to the tilt angle of the detected photon's path relative to the micro-objective optical axis (positive angles point away from the source). Five fiber-detector separations were considered: $1830$, $1730$ and $1910\um$ correspond to the middle, left, and right extremes of the $2.5\times$ micro-objective FOV, respectively, while $910$ and $3970\um$ correspond to positions roughly half and twice these distances, respectively. The median exit angle for each distribution is noted with a dashed line. A Lambertian exit angle distribution, characteristic of isotropic illumination, is also shown for comparison.
\label{fig:angularDist}}
\end{center}
\end{suppFigure}

\begin{suppVideo}
\begin{center}
\includegraphics{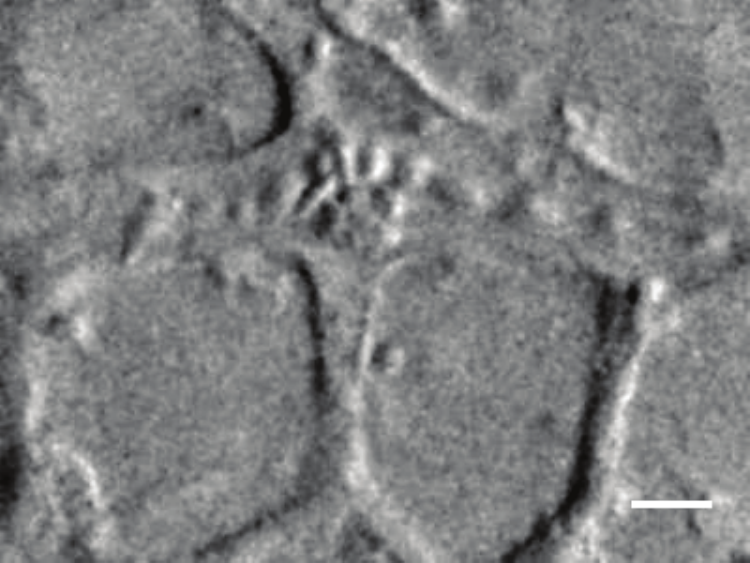}
\caption{CAM vasculature of day 11 chick embryo visualized with OBM. Capillary vessel walls are clearly visible, as are the dynamics of individual red blood cells. Axial resolution is made apparent by manually focusing up and down between the CAM mesoderm and ectoderm. Our imaging depth is limited here by the $60\um$ working distance of our micro-objective. Deeper imaging could be achieved with a longer working distance. In principle, maximum imaging depth of OBM is expected to be similar to that obtained with DIC, namely on the order of $l_s$. Scale bar $20\um$, imaging speed $17.5\Hz$.
\label{fig:capillaryVid}}
\end{center}
\end{suppVideo}

\begin{suppVideo}
\begin{center}
\includegraphics{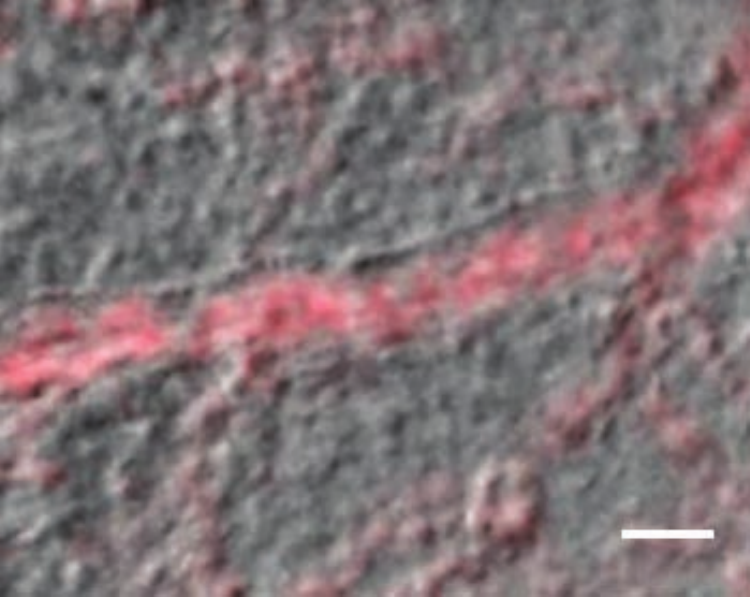}
\caption{OBM video of CAM vasculature of day 11 chick embryo. Moving RBCs are highlighted in red using a sliding 3-frame temporal variance filter. Scale bar $50\um$, imaging speed $17.5\Hz$.
\label{fig:red1XVid}}
\end{center}
\end{suppVideo}

\begin{suppVideo}
\begin{center}
\includegraphics{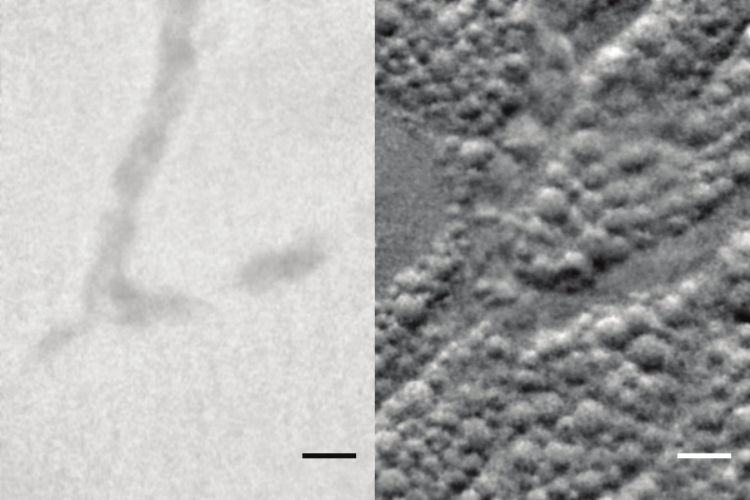}
\caption{Simultaneously acquired absorption and phase gradient OBM video of CAM vasculature of day 11 chick embryo. The probe was scanned over the sample using manually controlled translation stages. Scale bars $20\um$, imaging speed $17.5\Hz$. 
\label{fig:absphaVid}}
\end{center}
\end{suppVideo}

\begin{suppVideo}
\begin{center}
\includegraphics{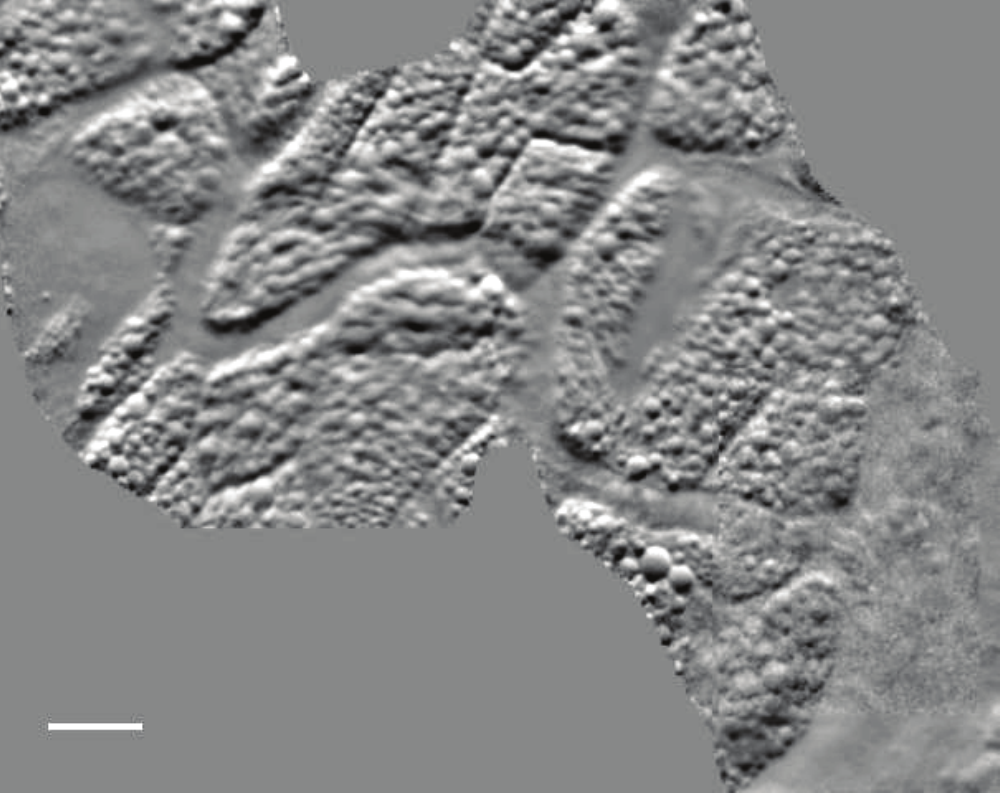}
\caption{Mosaic created from OBM video of CAM vasculature of day 11 chick embryo. Scale bar $50\um$, imaging speed $17.5\Hz$.
\label{fig:mosaicVid}}
\end{center}
\end{suppVideo}

\begin{suppVideo}
\begin{center}
\includegraphics{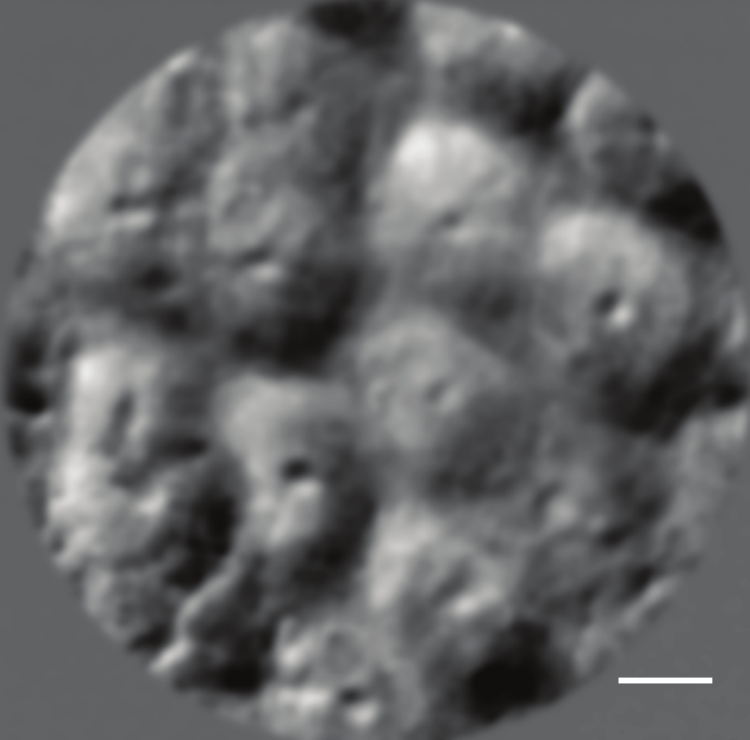}
\caption{OBM video of crypts of Lieberk\"{u}hn in excised mouse distal colon. Scale bar $30\um$, imaging speed $5\Hz$.
\label{fig:colonVid}}
\end{center}
\end{suppVideo}

\begin{suppVideo}
\begin{center}
\includegraphics{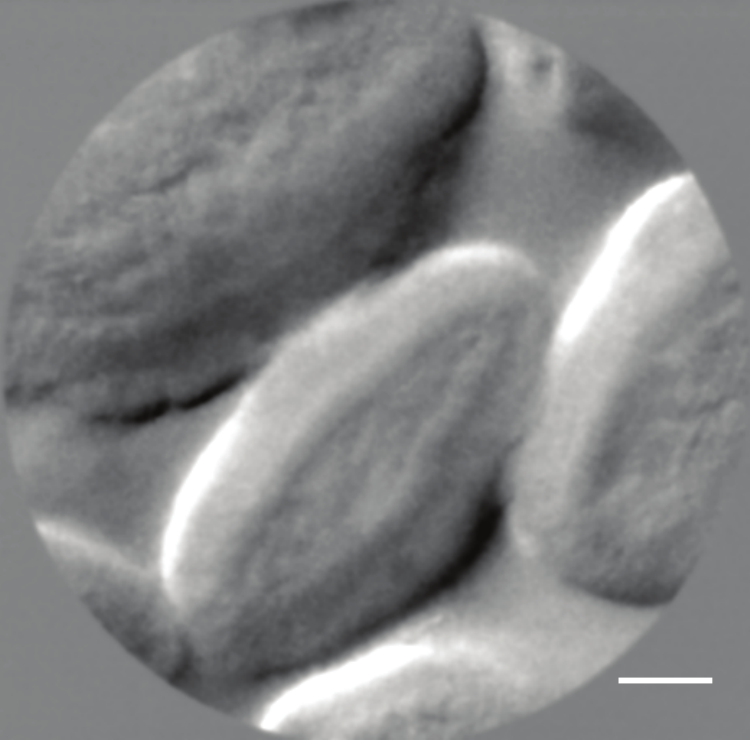}
\caption{OBM video of ileal villi in excised mouse small intestine. Scale bar $30\um$, imaging speed $5\Hz$.
\label{fig:ileumVid}}
\end{center}
\end{suppVideo}

\end{document}